\documentclass[12pt]{article}
\usepackage{epsf}
\usepackage{cite}

\begin{document}

\renewcommand{\topfraction}{0.8}
\renewcommand{\textfraction}{0.2}

\newcommand{\beq}{\begin{equation}}
\newcommand{\eeq}{\end{equation}}
\newcommand{\beqa}{\begin{eqnarray}}
\newcommand{\eeqa}{\end{eqnarray}}

\newcommand{\gsim}{\buildrel > \over {_\sim}}
\newcommand{\lsim}{\buildrel < \over {_\sim}}
\newcommand{\ie}{i.e.}
\newcommand{\eg}{e.g.}
\newcommand{\cf}{{\it cf.}}
\newcommand{\etal}{{\it et al.}}
\newcommand{\gev}{{\rm GeV}}
\newcommand{\jpsi}{J/\psi}
\newcommand{\ztodpj}{Z^0 \rightarrow {\rm direct \; prompt} \; J/\psi}
\newcommand{\order}[1]{${\cal O}(#1)$}
\newcommand{\eq}[1]{Eq.~(\ref{#1})}
\newcommand{\fig}[1]{Fig.~\ref{#1}}
\newcommand{\refcite}[1]{Ref.~\cite{#1}}
\newcommand{\Ariadne}{{\sc Ariadne}}
\renewcommand{\to}{\rightarrow}

\newcommand{\element}[1]{\langle 0 | {\cal O}_8^{#1}
\left( ^3S_1 \right) | 0 \rangle}

\newcommand{\PL}[3]{Phys.\ Lett.\ {#1} ({#3}) {#2}}
\newcommand{\NP}[3]{Nucl.\ Phys.\ {#1} ({#3}) {#2}}
\newcommand{\PRD}[3]{Phys.\ Rev.\ D {#1} ({#3}) {#2}}
\newcommand{\PRL}[3]{Phys.\ Rev.\ Lett.\ {#1} ({#3}) {#2}}
\newcommand{\ZPC}[3]{Z. Phys.\ C {#1} ({#3}) {#2}}
\newcommand{\PRe}[3]{Phys.\ Rep.\ {#1} ({#3}) {#2}}
\newcommand{\Comp}[3]{Comput.\ Phys.\ Comm.\ {#1} ({#3}) {#2}}
\newcommand{\ibid}[3]{{#1} ({#3}) {#2}}
\newcommand{\preprint}{Report No.\ }

\begin{titlepage}
\begin{flushright}
  NORDITA-96/78 P
\end{flushright}
\begin{flushright}
  hep-ph/9612408 \\ \today
\end{flushright}
\vskip .8cm
\begin{center}
  {\Large Evolution Effects in \\ $Z^0$ Fragmentation into Charmonium}
  \vskip .8cm
  {\bf P. Ernstr\"om, L. L\"onnblad and M. V\"anttinen}
  \vskip .5cm
  NORDITA\\
  Blegdamsvej 17, DK-2100 Copenhagen \O
  \vskip 1.8cm
\end{center}

\begin{abstract}

  \noindent
  In $Z^0$ decay into prompt charmonium, \ie\ charmonium not
  originating from $B$-meson decays, the most important contribution
  is expected to come from colour-octet mechanisms.  However, previous
  fixed-order calculations of the colour-octet contribution contain
  large logarithms which, in a more complete treatment, should be
  resummed to all orders. We study this resummation by using a Monte
  Carlo QCD cascade model and find that the fixed-order colour-octet
  result is diminished by 15\%. We compare the Monte Carlo
  calculations with results obtained by using analytical evolution
  equations.

\end{abstract}

\end{titlepage}

\setlength{\baselineskip}{7mm}

\section{Introduction}

The production of charmonium and bottomonium states in various
processes, especially at high-energy colliders \cite{BFY}, has
recently received considerable experimental and theoretical interest.
New data have become available from $p\bar{p}$ \cite{Tevatron}, $ep$
\cite{HERA} and $e^+e^-$ \cite{OPAL:upsilon,OPAL:psi,Delphi}
colliders. Theoretically, it has been realized that quarkonium
production at colliders is dominated by parton fragmentation
\cite{BraatenYuan}.  For the factorization of the hard dynamics of
$Q\bar{Q}$ production and the soft dynamics of quarkonium bound state
formation, a new framework is provided by non-relativistic QCD (NRQCD)
\cite{BBL}. This new theoretical approach has had initial successes
and is now approaching a stage where its predictive power can really
be judged by testing relations between various observables.

Within NRQCD, quarkonium is produced either through colour-singlet
mechanisms, where a perturbatively produced colour-singlet $Q\bar{Q}$
pair evolves non-perturbatively into a quarkonium state, or through
colour-octet mechanisms, where the intermediate $Q\bar{Q}$ pair is in
a colour-octet state and the non-perturbative transition involves the
exchange of soft gluons with the environment. Before the advent of
NRQCD, quarkonium production was usually calculated taking into
account only the colour-singlet channels. Contributions from
colour-octet channels are usually suppressed by powers of $v^2$, the
intrinsic velocity squared of the bound state, which apparently
justifies the colour-singlet model.  However, there are reactions
where this suppression is either absent or compensated by an
enhancement of colour-octet channels by powers of the strong coupling
constant $\alpha_s$.

A well-known example is the production of prompt $\psi'$ or direct
prompt $J/\psi$ in $p\bar{p}$ collisions at large $p_\perp$. (By
``prompt'' we mean those charmonium states which are not produced in
$B$-meson decays, and by ``direct'' those which are not produced in
the decays of other charmonium states.) The production rates measured
by the CDF Collaboration \cite{anomaly} are more than an order of
magnitude above the colour-singlet-model prediction. This anomaly can
be explained in the framework of NRQCD \cite{BraatenFleming}, which
includes a colour-octet production channel whose contribution is
proportional to the probability of a non-perturbative transition, to
be treated as a free parameter of the theory and fitted to the CDF
data.

The fact that NRQCD comes with a number of free parameters (although
there are scaling rules which determine the order of magnitude of the
transition probabilities in terms of $v^2$) necessitates cross checks
between different experiments.  Colour-octet contributions to \eg\ 
photoproduction \cite{photo} and fixed-target hadroproduction
\cite{fixed} have indeed been calculated, partly in order to provide
such cross checks, partly as an attempt to explain discrepancies
between the colour singlet model and experimental data \cite{VHBT}.

An interesting laboratory of quarkonium production is provided by
$e^+e^-$ annihilation at the $Z^0$ pole. Experimental branching ratios
or upper limits for $Z^0$ decay into prompt quarkonium have recently
been published by the OPAL \cite{OPAL:upsilon,OPAL:psi} and DELPHI
\cite{Delphi} Collaborations. Theoretical calculations
\cite{BraatenYuan,Kuhn,Hagiwara,BCY,Barger,CKY,Cho} show that the
largest contribution is expected to come from the same
non-perturbative transition that in the NRQCD framework explains the
$\psi'$ anomaly. Using the value of the NRQCD matrix element from
Tevatron fits, the magnitude of this contribution is consistent with
the experimental numbers \cite{CKY,Cho}. However, the existing
calculations have been done at a fixed order in $\alpha_s$, and the
result includes large logarithms, $\ln(M_Z^2/M_{J/\psi}^2)$, which
call for a resummation.  It is the purpose of this paper to do this
resummation and see how much the fixed-order results are changed.

Our principal tool in studying the evolution effects is a Monte Carlo
QCD cascade model, implemented in the \Ariadne\ program
\cite{program}.  This approach was used in earlier work
\cite{Nordita9640} on the large $p_\perp$ hadroproduction of
quarkonium.

The evolution effects can also be studied using analytical evolution
equations \cite{modified} to determine the scale dependence of
fragmentation functions.  Since the analytic fragmentation formalism
fails in the threshold region where the colour-octet fragmentation
process peaks, we derive our final results using the Monte Carlo
method.  Above the threshold region we use the analytical evolution
equations as a check of the Monte Carlo implementation.

In Section \ref{mechanisms} below, we discuss in detail the
colour-singlet and colour-octet mechanisms of parton fragmentation
into charmonium and review the earlier work of other authors.  We also
describe the analytical fragmentation formalism used as a check of our
Monte Carlo simulation.  In Section \ref{MC} the Monte Carlo
simulation of evolution effects with the \Ariadne\ program is
discussed.  The Monte Carlo method is then applied to the colour-octet
fragmentation process to obtain our main results. These are presented
in Section \ref{results}, where we show that the evolution effects
diminish the theoretical prediction of $Z^0$ branching ratio into
prompt charmonium by about 15\%. Finally, the results are discussed in
Section \ref{discussion}.

\section{QCD production mechanisms \label{mechanisms}}

In $Z^0$ decay into $^3S_1$ charmonium (which we denote generically as
$\psi$) the dominant colour-singlet contribution comes from charm
fragmentation, \ie\ the process where $Z^0$ decay into $c\bar c$ is
followed by the perturbative QCD process $c \to c\bar c \left[
  ^3S_1^{(1)} \right] + c$ and then the non-perturbative process
\mbox{$c\bar c \left[ ^3S_1^{(1)} \right] \to \psi$}.  There is an
equal contribution from $\bar c$ fragmentation.  The cross section is
\begin{equation}
  \frac{d\Gamma}{dz}(Z^0 \to \psi + X)
  = 2 \, \Gamma(Z^0 \to c\bar c) \, D_c^\psi(z,\mu),
  \label{cprodXcfrag}
\end{equation}
where $z = E_\psi/(M_Z/2)$ is the fraction of the initial charm quark
energy taken by the $\psi$, $D_c^\psi(z,\mu)$ is the charm
fragmentation function into $\psi$, and $\mu = O(M_Z/2)$ is the
fragmentation scale \cite{BCY}.  The normalization of this
contribution is related to the leptonic width of the $\psi$.

In \cite{BCY} large logarithms $\ln(M_Z^2/M_{J/\psi}^2)$ were resummed
using Altarelli-Parisi evolution equations \cite{AP} for the scale
dependence of $D_c^\psi(z,\mu)$.  Here we will use both Monte Carlo
simulations and improved analytical evolution equations
\cite{modified} to study the evolution effects.

The dominant colour-octet contribution comes from fragmentation
mechanisms where $Z^0$ decay into $q\bar q g$ is followed by the
perturbative splitting $g \to c\bar c\left[ ^3S_1^{(8)} \right]$ and
then the non-perturbative process $c\bar c\left[ ^3S_1^{(8)} \right]
\to \psi + X$.  A leading contribution to the non-perturbative process
comes from the emission of two soft gluons. The probability of this
non-perturbative transition is proportional to the NRQCD matrix
element $\element{\psi}$, \ie\ the same element whose value has been
fit to reproduce the CDF data. The normalization of the colour-octet
component can therefore be obtained from Tevatron fits.

In the following, all numbers were derived using the colour-octet
matrix element $\element{J/\psi} = 0.0066 \; ({\rm GeV})^3$ obtained
from a fit \cite{ChoLeibovich} of the CDF data and the colour-singlet
matrix element $\langle 0 | {\cal O}_1^{J/\psi}(^3S_1) | 0 \rangle =
0.73 \; ({\rm GeV})^3$ \cite{BCY}, related to the wavefunction at the
origin of coordinate space $R_S(0)$ through $|R_S(0)|^2= (2\pi/9)
\langle 0 | {\cal O}_1^{J/\psi}(^3S_1) | 0 \rangle = 0.512 \; ({\rm
  GeV})^3$.  Furthermore, $m_c = 1.5$ GeV, $M_\psi = 2m_c$, and
$\alpha_s(2m_c) = 0.253$ \cite{CKY} were used.

Below, we shall speak of $Z^0$ decay into direct prompt $J/\psi$.  The
production of $\chi_c$ and $\psi'$ from $c\bar c\left[ ^3S_1^{(8)}
\right]$ intermediate states and their decays into $J/\psi$ can be
included by multiplying the octet contribution by a factor 2.1,
derived from the octet matrix elements of Ref.\ \cite{ChoLeibovich}
and the branching ratios given in \cite{PDG}. The production of
$\psi'$ from $c\bar c\left[ ^3S_1^{(1)} \right]$ and its decay into
$J/\psi$ can be included by multiplying the singlet contribution by a
factor 1.3, derived using the colour-singlet-model relation
\begin{equation}
  \frac{|R_{2S}(0)|^2}{|R_{1S}(0)|^2}
  = \left( \frac{M_\psi'}{M_{J/\psi}} \right)^2
    \; \frac{\Gamma(\psi' \to \ell \ell)}{\Gamma(J/\psi \to \ell \ell)}
\end{equation}
with experimental leptonic widths \cite{PDG}.

\begin{figure}[htbp]
\begin{center}
\leavevmode
%{\epsfxsize=10truecm \epsfbox{/home/mv/Paperit/NORDITA-96-XXP/Kuvat/fig1.ps}}
{\epsfxsize=10truecm \epsfbox{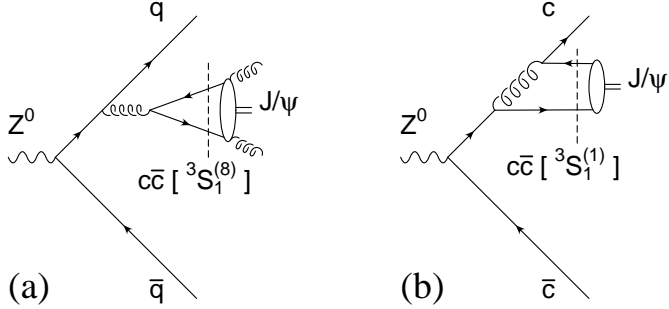}}
\end{center}
\caption[*]{{\it The Feynman diagrams which describe
    the dominant colour-octet and colour-singlet mechanisms of $Z^0$
    fragmentation into prompt $J/\psi$. (a) Colour-octet gluon
    fragmentation. There is another diagram where the gluon is emitted
    by the antiquark line. (b) Colour-singlet charm fragmentation.
    There is (in an axial gauge) another diagram which represents
    $\bar c$ fragmentation.}}
\label{Fig:diagrams}
\end{figure}

This far, the calculation of the colour-octet contribution has only
been done at a fixed order in perturbative QCD.  The process is then
represented by the Feynman diagram shown in \fig{Fig:diagrams}a and
another diagram where the gluon is emitted by the antiquark line. The
resulting charmonium energy spectrum is \cite{CKY,Cho}
\begin{eqnarray}
  \lefteqn{\frac{d\Gamma}{dz} (Z^0 \to \psi + X)
  = \frac{\alpha_s^2(2m_c)}{18} \, \Gamma(Z^0 \to q\bar{q}) \,
    \frac{\element{\psi}}{m_c^3}}
    \nonumber \\
  & \times & \left[ \rule{0mm}{8mm} \left( \frac{1 + (1-z)^2}{z}
      + 2\left(\frac{M_\psi}{M_Z}\right)^2 \frac{2-z}{z}
      + \left(\frac{M_\psi}{M_Z}\right)^4 \frac{2}{z} \right)
       \right.
      \nonumber \\
  & & \times \left. \mbox{}
      \ln \left( \frac{z + \sqrt{z^2 - 4(M_\psi/M_Z)^2}
      }{z - \sqrt{z^2 - 4(M_\psi/M_Z)^2}} \right)
      - 2 \sqrt{z^2 - 4(M_\psi/M_Z)^2} \rule{0mm}{8mm} \right] ,
  \label{CKYresult}
\end{eqnarray}
which leads to an integrated width
\begin{eqnarray}
\lefteqn{\Gamma(Z^0 \to \psi+X) = \Gamma(Z^0 \to q\bar{q}) \cdot
     { \alpha_s^2(2m_c) \,
       \element{\psi}  \over 36 m_c^3 }}\nonumber \\
    & \times & \left[ 
         5 - { \pi^2 \over 3 } + 3\log\left( M_\psi^2 \over M_Z^2 \right) +
         \log^2\left(M_\psi^2 \over M_Z^2\right) +
         {\cal O}\left(\frac{M_\psi^2}{M_Z^2}\right)
     \right],
  \label{CKYintegratedResult}
\end{eqnarray}
giving \mbox{${\rm Br}^{(8)}(\ztodpj) = 0.68 \cdot 10^{-4}$} for the
$J/\psi$.  Hence this contribution is clearly larger than the
colour-singlet charm fragmentation contribution \mbox{${\rm
    Br}^{(1)}(\ztodpj) = 0.22 \cdot 10^{-4}$} \cite{BCY}.  The $z$
distribution is also markedly different, peaking around \mbox{$z=0.1$}
whereas the singlet component peaks around \mbox{$z=0.8$}.

Far from the threshold ($ z \gg M_\psi / M_Z $) the differential decay
width can be written as
\begin{eqnarray}
  \frac{d\Gamma}{dz} & = &
  \frac{\alpha_s^2(2m_c)}{18} \Gamma(Z^0 \to q\bar{q})
  \frac{\langle 0 | {\cal O}_8^\psi(^3S_1) | 0 \rangle}{m_c^3}\nonumber\\
  & &\times\left[ \frac{1 + (1-z)^2}{z}
      \ln \left( \frac{z^2}{M_\psi^2/M_Z^2} \right) - 2z
      + {\cal O}(\frac{M_\psi^2}{z^2M_Z^2}) \right].
  \label{CKYresultLimit}
\end{eqnarray}
The large logarithm arises from the nearly collinear splitting of an
initial quark into a quark and a virtual gluon, the virtuality
$M_\psi^2$ being small on the scale set by the $Z^0$ mass. The double
logarithm in \eq{CKYintegratedResult} arises from the soft gluon
emission represented by the factors $1/z$ in \eq{CKYresult}.  In a
complete calculation such logarithms should be resummed to all orders
in perturbation theory.

Just as in the colour singlet charm fragmentation case, such a
resummation can be performed by rewriting the decay width in terms of
fragmentation functions and then resumming the large logarithms using
analytical evolution equations for the fragmentation functions.  Such
a treatment does, however, neglect terms of ${\cal O}(M_\psi^2/(z^2
M_Z^2))$, which become important in the threshold region, where the
colour octet $J/\psi$--production peaks.  We will therefore perform
the resummation by means of a Monte Carlo simulation, and use the
analytical resummation described below only as check at large $z$.
Note that even at the threshold the virtuality of the fragmenting
quark can be as low as of the order $M_\psi M_Z$ which is still much
smaller than $M_Z^2$, thus motivating a fragmentation approach.

In fragmentation language, the Feynman diagram in \fig{Fig:diagrams}a
may be viewed either as quark production followed by quark
fragmentation into $J/\psi$ (via splitting), or if the quark--gluon
invariant mass is large, as gluon production followed by gluon
fragmentation into $J/\psi$.  Correspondingly, the colour octet
contribution can be written as \cite{BFY,CKY,Fleming}
\begin{eqnarray}
  { d\Gamma \over dz }(Z^0 \to \psi + X) & = &
  2 \, \Gamma( Z^0 \to q \bar{q} )D_q^\psi(z,\mu^2) \nonumber\\
 & & + \int_0^1 {dx \over x}
  {d\Gamma( Z^0 \to q \bar{q} g ) \over dz_g}(z/x,\mu^2)
  D_g^\psi(x,\mu^2) ,
\label{TwoTermFragmentation}
\end{eqnarray}
where $\mu$ sets the scale for the fragmentation functions and defines
the minimal $qg$ or $\bar{q}g$ invariant mass for the differential
three-parton decay width
\begin{eqnarray}
  \lefteqn{{d\Gamma( Z^0 \to q \bar{q} g ) \over dz_g}(z_g,\mu^2) =
  { C_F \alpha_s(\mu^2) \over \pi }\Gamma( Z^0 \to q \bar{q} )}\nonumber\\
 & & \times
  \left(
    \frac{1+(1-z)^2}{z}\log \left( z_g-\mu^2/M_Z^2 \over \mu^2/M_Z^2 \right)
     - z_g + { 2 \mu^2 \over M_Z^2 }
  \right) .
\label{gProduction}
\end{eqnarray}
Inserting LO expressions for the fragmentation functions
\beqa
  D_q^\psi(z,\mu^2) & = & 
    \frac{\alpha_s^2 \, \langle 0 | {\cal O}_8^\psi(^3S_1) | 0 \rangle}
         {36 m_c^3 }
    \left(
       \frac{1+(1-z)^2}{z}\log \left( z \mu^2 \over M_\psi^2 \right)
       - z + { M_\psi^2 \over \mu^2  }
    \right)
\\
  D_g^\psi(z,\mu^2) & = & 
    \frac{\pi \, \alpha_s \, \langle 0 | {\cal O}_8^\psi(^3S_1) | 0 \rangle}
         {24 m_c^3 }
    \, \delta(1-z)\label{LOoctetFrfn}
\eeqa
in \eq{TwoTermFragmentation} the LO large $z$ result
\eq{CKYresultLimit} is reproduced up to corrections
\order{\mu^2/M_Z^2}.

At scales $\mu$ of the order of $M_Z$, \eq{TwoTermFragmentation} is
totally dominated by the quark fragmentation term.  The corrections
are \order{1}, but the leading logarithmic terms are still correctly
reproduced.

To resum the large logarithms we have used the improved evolution
equations \cite{modified}
\beqa
  \mu^2 {\partial \over \partial \mu^2} D^\psi_i(z,\mu^2) & \! = \! &
    d^\psi_i(z,\mu^2) +
    { \alpha_s(\mu^2) \over 2 \pi }
    \sum_j  \int_z^1 {dy \over y}
       D^\psi_j(z/y,y\mu^2)
       P_{j i}(y)
\label{improvedEvolutionEq}
   \\
   D^\psi_i(z,M_\psi^2/z)  & \! = \! & 0 .
\eeqa
In the colour singlet case we have used the following charm
fragmentation source term extracted from \cite{BCY}
\beqa
 d_c^\psi(z,x m_c^2) &=&
    { 8 \alpha_s^2 |R(0)|^2 \over  27 \pi m_c^3 }
    { x \over (x-1)^4 }
    \biggl[
      (x^2 - 2x - 47) - z(x-1)(x-9) + 
\nonumber
\\ & & 
     + { 4z(1-z) \over 2-z }x(x-1) - { 4(8-7z-5z^2)\over (2-z)}(x-1) +
\nonumber
\\ & & 
      + { 12z^2 (1-z)\over(2-z)^2}(x-1)^2
    \biggr]
    \theta\left( x- { 4 \over z } - { 1 \over 1-z } \right).
\label{dc}
\eeqa
Using this source term we have solved the evolution equation
(\ref{improvedEvolutionEq}) numerically and used \eq{cprodXcfrag} to
find the colour singlet contribution to $J/\psi$ production in $Z^0$
decays.

In the colour octet case the gluon fragmentation source term is a
double delta function
\beq
  d_g^\psi(z,s)  = 
  \frac{\pi \, \alpha_s \, \langle 0 | {\cal O}_8^\psi(^3S_1) | 0 \rangle}
  {24 m_c^3 }
  \, \delta(1-z) \, \delta\left(1-{s \over M_\psi^2}\right).
  \label{doubledelta}
\eeq
Equivalently, one may omit this source term and use \eq{LOoctetFrfn}
as an initial condition at the scale $2m_c$.

%The LO colour octet $q \to q + \psi$ diagram may be viewed as a quark
%splitting into a gluon followed by gluon fragmentation into the
%$J/\psi$, but after subtracting the splitting contribution, a negative
%contribution to the quark fragmentation source term remains
%%
%\beq
%  d_q^\psi(z,s) = - 
%    \frac{ \alpha_s \, \langle 0 | {\cal O}_8^\psi(^3S_1) | 0 \rangle}
%         {36 m_c^3 }
%    \, { M_\psi^2 \over s }\, \theta(s-\frac{M_\psi^2}{z}),
%\eeq
%%
%necessary to reproduce the LO quark fragmentation function but without
%numerical importance.

We have solved the evolution equation (\ref{improvedEvolutionEq})
numerically and used \eq{TwoTermFragmentation}, performing the
convolution of the evolved gluon fragmentation function and the $Z^0
\to q \bar{q} g$ differential decay width \eq{gProduction}, to find
the colour octet contribution to $J/\psi$ production in $Z^0$ decays.

The commonly used AP evolution differs from the improved evolution of
\eq{improvedEvolutionEq} in two ways:
\begin{itemize}
\item In the AP evolution equations the source term
  $d^\psi_i(z,\mu^2)$ is put to zero and instead unphysical initial
  conditions are chosen at the $M_\psi$ scale such that the
  AP--evolved fragmentation functions agree with perturbative
  fragmentation functions at {\em large} scales.
%  When the improved
%  evolution equation is used, the fragmentation functions agree with
%  perturbative ones also at low scales.
\item In the AP evolution equations, the fragmentation functions on
  the right hand side of \eq{improvedEvolutionEq} are evaluated at the
  scale $\mu^2$ instead of at the scale $y \mu^2$, and thus, fail to
  take into account the momentum constraint saying that the maximum
  virtuality of a parton taking a positive light-cone momentum
  fraction $y$ from a parent parton with virtuality $\mu^2$, is $y
  \mu^2$.
\end{itemize}
Neither of these points is crucial for our comparison with the Monte
Carlo simulation at large scales and far from the threshold.

\section{The Monte Carlo method \label{MC}}

The implementation of charmonium production in the \Ariadne\ event
generator was described in \refcite{Nordita9640}, and here we will
only give the main points and present some improvements.

Three mechanisms are included (see below). In all cases, the
perturbative splitting into a $c\bar{c}$ pair and the non-perturbative
formation of the charmonium state are treated together as one step in
the otherwise purely perturbative QCD shower. The motivation behind
including the soft formation of a charmonium state in the perturbative
level, before the normal hadronization, is that all hard interactions
in principle are already included in the splitting function.

The three implemented mechanisms are treated somewhat differently:
\begin{itemize}
\item {\bf Colour octet gluon fragmentation:} implemented as a
  perturbative splitting of a gluon into a collinear colour octet
  $c\bar c \left[ ^3S_1^{(8)}\right]$ pair, followed by a
  non-perturbative transition into a $\psi$ and one soft gluon. To
  correctly conserve quantum numbers, there should be at least two
  gluons emitted, but since the subsequent string fragmentation is
  stable w.r.t.\ addition of soft gluons, emitting only one to
  conserve colour should give a good description of the final state
  hadrons.  For practical reasons, the delta function in the splitting
  kernel \mbox{$\delta(z-1)$} is given a small width by replacing it
  with step functions
  \mbox{$(1/\epsilon)\,\theta(z-(1-\epsilon))\,\theta(1-z)$.} It is
  now possible to study the final-state properties of the octet
  mechanism as a function of this width which typically is expected to
  be on the order of the squared velocity $v^2$ of the quarks inside
  the charmonium.
\item {\bf Colour singlet gluon fragmentation:} a gluon is
  perturbatively split into a collinear colour singlet $c\bar c \left[
    ^3S_1^{(8)}\right]$ pair and two hard gluons.  The $c\bar c$ pair
  directly forms a $\psi$ while the two gluons may continue cascading
  until hadronization sets in. In \refcite{Nordita9640} only one gluon
  was emitted to conserve colour and momentum, but now the full
  three-body phase space is explored with the two-gluon invariant mass
  correctly distributed according to \refcite{ColSingGlue}. The
  final-state distributions discussed in \refcite{Nordita9640} turned
  out to be insensitive to this change, and the conclusions therein
  still hold.  The colour singlet gluon fragmentation contribution to
  $J/\psi$ production at LEP is very small and will not be discussed
  further in this report.
\item {\bf Colour singlet charm quark fragmentation:} a charm quark is
  split into a collinear $c\bar c \left[ ^3S_1^{(8)}\right]$ pair and
  a charm quark according to Fig.~\ref{Fig:diagrams}b.  The $c\bar c$
  pair directly forms a $\psi$ while the charm quark may continue
  radiating.
\end{itemize}
Besides the treatment of the soft gluons in the octet model, the main
uncertainty in the implementation comes from the fact that the dipole
cascade model in the \Ariadne\ program in each step has all partons
on-shell. In order for a gluon to split into a charmonium, some energy
has to be {\it borrowed} from one of the colour-connected partons. As
discussed in \refcite{Nordita9640}, this is a safe procedure as long
as the cascade is strongly ordered, \ie\ when the transverse momentum
$p_{\perp g}$ of the gluon to be split is much larger than the
transverse mass $m_{\perp \psi}$ of the charmonium.  For
high-$p_\perp$ charmonium production at hadron colliders this is a
good approximation. But in $Z^0$ decays there may even be situations
where the mass $m_{\psi+g}$ of the formed charmonium-gluon state is
larger than $p_{\perp g}$, especially for the octet channel at small
$z$. We have tried to estimate these uncertainties by \eg\ explicitly
requiring $m_{\psi+g}<p_{\perp g}$, and the results turned out to be
insensitive to such modifications. Therefore we assume that the
description is good also at small $z$, although further studies are
needed.

In this paper we will only use this program to study $J/\psi$
production at LEP, but the implementation is general enough for
studying the fragmentation production of any quarkonia at any
experiment.

\section{Results \label{results}}

In this section we present the results of our Monte Carlo simulations.
We first compare with the analytical evolution, then we discuss the
energy isolation cuts of the experimental analysis in the presence of
evolution effects and finally proceed to determine the branching ratio
of $Z^0$ into prompt $J/\psi$.

In each case, we also discuss the dependence of our results on the
energy carried by the soft gluons emitted in the non-perturbative
colour-octet transitions. This is of the order of the quark kinetic
energies $m_c v^2/2$ in the charmonium rest frame and hence negligible
in the first approximation.

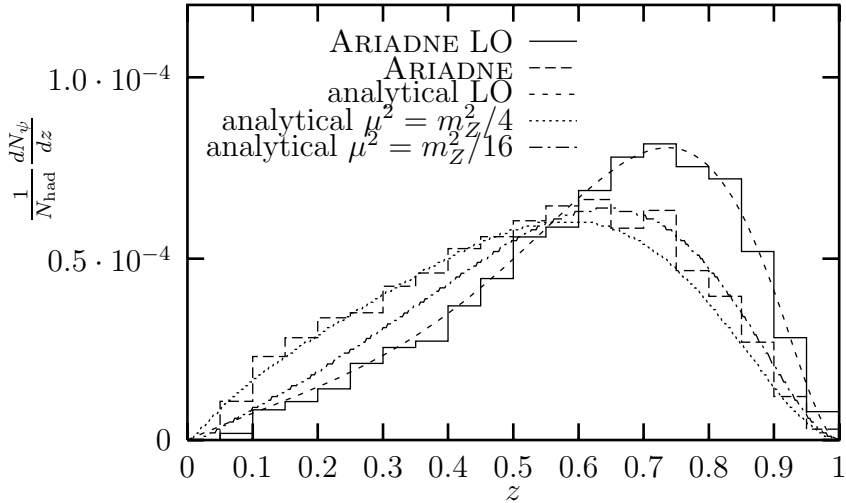
\begin{figure}[t]
%\input /home/leif/MCEG/Projects/Onium/Z0paper/czdist.tex
% GNUPLOT: LaTeX picture with Postscript
\setlength{\unitlength}{0.1bp}
\begin{picture}(3239,1943)(0,0)
\put(1828,1355){\makebox(0,0)[r]{analytical $\mu^2=m_Z^2/16$}}
\put(1828,1455){\makebox(0,0)[r]{analytical $\mu^2=m_Z^2/4$}}
\put(1828,1555){\makebox(0,0)[r]{analytical LO}}
\put(1828,1655){\makebox(0,0)[r]{\Ariadne}}
\put(1828,1755){\makebox(0,0)[r]{\Ariadne\ LO}}
\put(1828,51){\makebox(0,0){$z$}}
\put(100,1271){%
%
\makebox(0,0)[b]{\shortstack{$\frac{1}{N_{\mbox{\tiny had}}} \frac{dN_\psi}{dz}$}}%
%
}
\put(3056,151){\makebox(0,0){1}}
\put(2810,151){\makebox(0,0){0.9}}
\put(2565,151){\makebox(0,0){0.8}}
\put(2319,151){\makebox(0,0){0.7}}
\put(2074,151){\makebox(0,0){0.6}}
\put(1828,151){\makebox(0,0){0.5}}
\put(1582,151){\makebox(0,0){0.4}}
\put(1337,151){\makebox(0,0){0.3}}
\put(1091,151){\makebox(0,0){0.2}}
\put(846,151){\makebox(0,0){0.1}}
\put(600,151){\makebox(0,0){0}}
\put(540,1619){\makebox(0,0)[r]{$1.0\cdot 10^{-4}$}}
\put(540,935){\makebox(0,0)[r]{$0.5\cdot 10^{-4}$}}
\put(540,251){\makebox(0,0)[r]{$0$}}
\end{picture}
\caption[dummy]{{\it The $z$ distribution of $J/\psi$ from
    colour-singlet charm quark fragmentation. Solid and dashed
    histograms are from \Ariadne\ in leading order and with full
    evolution respectively. The dashed line is from an analytical
    calculation to leading order, and the dotted and dash-dotted line
    are from analytical calculations with improved evolution using two
    different scales.}}
\label{Fig:Dc}
\end{figure}

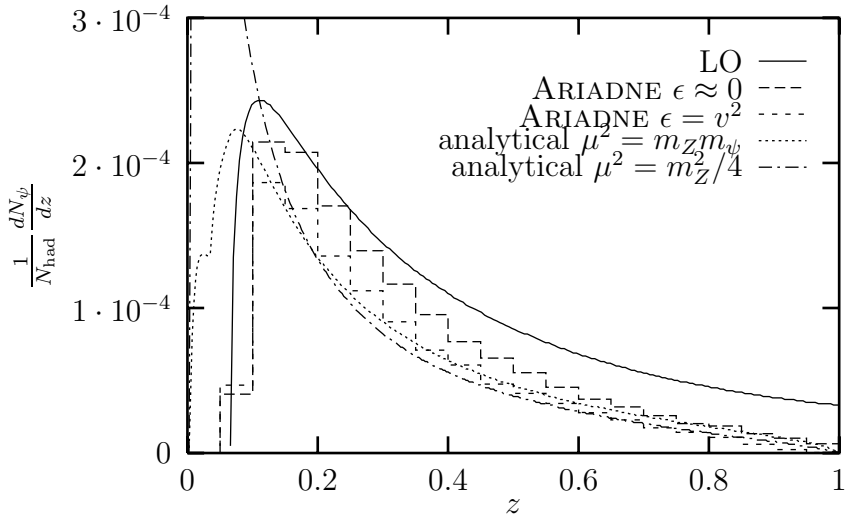
\begin{figure}[t]
%\input /home/leif/MCEG/Projects/Onium/Z0paper/ozdist.tex
% GNUPLOT: LaTeX picture with Postscript
\setlength{\unitlength}{0.1bp}
\begin{picture}(3239,1943)(0,0)
\put(2693,1329){\makebox(0,0)[r]{analytical $\mu^2=m_Z^2/4$}}
\put(2693,1429){\makebox(0,0)[r]{analytical $\mu^2=m_Zm_\psi$}}
\put(2693,1529){\makebox(0,0)[r]{\Ariadne\ $\epsilon=v^2$}}
\put(2693,1629){\makebox(0,0)[r]{\Ariadne\ $\epsilon\approx 0$}}
\put(2693,1729){\makebox(0,0)[r]{LO}}
\put(1828,51){\makebox(0,0){$z$}}
\put(100,1071){%
%
\makebox(0,0)[b]{\shortstack{$\frac{1}{N_{\mbox{\tiny had}}} \frac{dN_\psi}{dz}$}}%
%
}
\put(3056,151){\makebox(0,0){1}}
\put(2565,151){\makebox(0,0){0.8}}
\put(2074,151){\makebox(0,0){0.6}}
\put(1582,151){\makebox(0,0){0.4}}
\put(1091,151){\makebox(0,0){0.2}}
\put(600,151){\makebox(0,0){0}}
\put(540,1892){\makebox(0,0)[r]{$3\cdot 10^{-4}$}}
\put(540,1345){\makebox(0,0)[r]{$2\cdot 10^{-4}$}}
\put(540,798){\makebox(0,0)[r]{$1\cdot 10^{-4}$}}
\put(540,251){\makebox(0,0)[r]{0}}
\end{picture}
\caption[dummy]{{\it The $z$ distribution of $J/\psi$ from
    colour-octet gluon fragmentation. The solid line is from an
    analytical calculation to leading order. The dotted and
    dash-dotted lines are from analytical calculations with evolution
    using two different scales. The histograms are from \Ariadne\ 
    using different widths of the step function in the splitting
    function: long-dashed corresponds almost to a delta function
    ($\epsilon=10^{-4}$), and short-dashed to a width of
    \mbox{$\epsilon=0.2\approx v^2$}.}}
\label{Fig:Do}
\end{figure}

In Figs.\ \ref{Fig:Dc} and \ref{Fig:Do} we compare the results using
the analytical fragmentation formalism with the results from the Monte
Carlo simulation.  The results agree well in the large $z$ limit.
Closer to the threshold region, where the analytical fragmentation
formalism is known to fail, the disagreement becomes large, especially
for octet gluon fragmentation.  If we introduce a finite width in
\eq{doubledelta}, corresponding to an energy of the order of $M_\psi
v^2$ carried by the soft gluons in the non-perturbative transition
$c\bar c \left[ ^3S_1^{(8)} \right] \to \psi + gg$, the distribution
somewhat softened, as expected.

In the experimental analysis, the separation of prompt $J/\psi$ from
the $B$-decay component is primarily based on energy isolation cuts,
supplemented by vertex detector information \cite{OPAL:psi,Delphi}.
The $J/\psi$ produced in $B$-meson decays are typically accompanied by
approximately collinear, energetic light hadrons. The mechanisms of
parton fragmentation into prompt $J/\psi$, on the other hand, have a
tendency to kick the $c\bar c$ pair rather far apart from the original
parton direction, resulting in the production of charmonia with little
accompanying energy. Prompt charmonia can therefore be identified by
requiring the accompanying energy $E_{\rm cone}$ within a cone of
half-opening angle $\alpha$ around the charmonium direction to be less
than some $E_0$. The LEP groups use $E_0 = 4$ GeV and $\alpha =
20^\circ$ \cite{Delphi} or $30^\circ$ \cite{OPAL:psi}.

We have studied how much the evolution effects change the distribution
of energy around the charmonium. The results are presented in Figs.\ 
\ref{Fig:cut} and \ref{Fig:ocut}, where we plot the distribution of
events as a function of accompanying energy, $(d\Gamma/dE_{\rm
  cone})/\Gamma$.  In the majority of events, charmonium states coming
from $B$ decays are accompanied with energy $E_{\rm cone} > 4$ GeV;
the prompt component, on the other hand, contains a dominant peak at
$E_{\rm cone} \simeq 0$, representing events where there are no hard
partons within $30^\circ$ of the charmonium momentum direction.
Although the fixed-order result for prompt production is changed by
including the effects of the parton cascade, hadronization and a
finite energy of the soft gluons emitted in the octet channel, the
fixed-order calculation already gives a good idea of how the isolation
cuts work in a realistic situation. Hence we do not expect any large
corrections to the results of Refs.\ \cite{OPAL:psi} and
\cite{Delphi}.

\begin{figure}[t]
%\input /home/leif/MCEG/Projects/Onium/Z0paper/cedist.tex
% GNUPLOT: LaTeX picture with Postscript
\setlength{\unitlength}{0.1bp}
\begin{picture}(3239,1943)(0,0)
\put(2442,1364){\makebox(0,0)[r]{B decays}}
\put(2442,1464){\makebox(0,0)[r]{Full \Ariadne}}
\put(2442,1564){\makebox(0,0)[r]{\Ariadne\ LO}}
\put(1908,51){\makebox(0,0){$E_{\mbox{\tiny cone}}$ (GeV)}}
\put(100,1071){%
%
\makebox(0,0)[b]{\shortstack{$\frac{1}{N_\psi} \frac{dN_\psi}{dE_{\mbox{\tiny cone}}}$}}%
%
}
\put(3056,151){\makebox(0,0){40}}
\put(2442,151){\makebox(0,0){30}}
\put(1828,151){\makebox(0,0){20}}
\put(1214,151){\makebox(0,0){10}}
\put(600,151){\makebox(0,0){0}}
\put(540,1892){\makebox(0,0)[r]{0.15}}
\put(540,1345){\makebox(0,0)[r]{0.1}}
\put(540,798){\makebox(0,0)[r]{0.05}}
\put(540,251){\makebox(0,0)[r]{0}}
\end{picture}
\caption[dummy]{{\it The distribution of $J/\psi$ events from
    charm fragmentation and from $B$ decays as a function of the
    accompanying energy $E_{\rm cone}$ within a cone of half-opening
    angle $\alpha = 30^\circ$ around the $J/\psi$ direction.  Solid
    histogram is charm fragmentation from \Ariadne\ to leading order
    on the parton level. Long-dashed histogram is the same but with
    full evolution and hadronization. Short-dashed histogram is
    $J/\psi$ from B decays. The distributions have been normalized to
    unity.}}
\label{Fig:cut}
\end{figure}
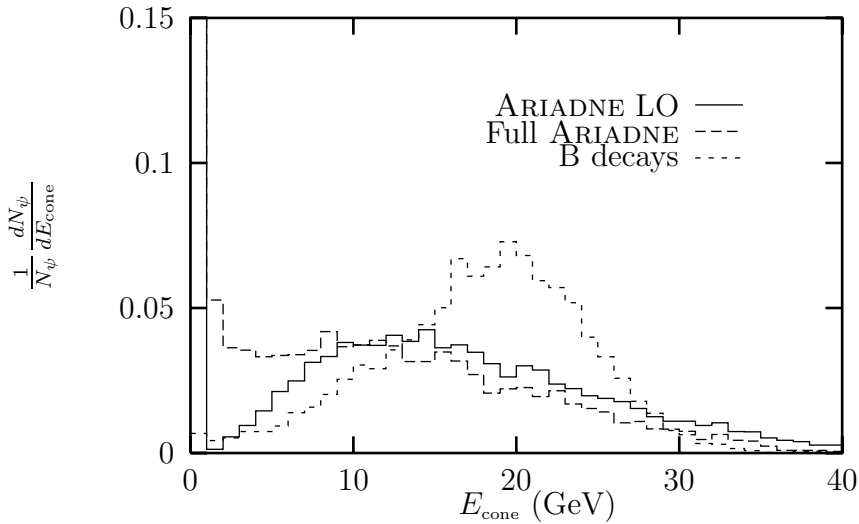

\begin{figure}[t]
%\input /home/leif/MCEG/Projects/Onium/Z0paper/oedist.tex
% GNUPLOT: LaTeX picture with Postscript
\setlength{\unitlength}{0.1bp}
\begin{picture}(3239,1943)(0,0)
\put(2275,1264){\makebox(0,0)[r]{analytical LO $\epsilon=0$}}
\put(2275,1364){\makebox(0,0)[r]{\Ariadne\ $\epsilon=v^2$}}
\put(2275,1464){\makebox(0,0)[r]{\Ariadne\ $\epsilon\approx 0$}}
\put(2275,1564){\makebox(0,0)[r]{\Ariadne\ LO $\epsilon\approx 0$}}
\put(1828,51){\makebox(0,0){$E_{\mbox{\tiny cone}}$ (GeV)}}
\put(100,1071){%
%
\makebox(0,0)[b]{\shortstack{$\frac{1}{N_\psi} \frac{dN_\psi}{dE_{\mbox{\tiny cone}}}$}}%
%
}
\put(2833,151){\makebox(0,0){40}}
\put(2275,151){\makebox(0,0){30}}
\put(1716,151){\makebox(0,0){20}}
\put(1158,151){\makebox(0,0){10}}
\put(600,151){\makebox(0,0){0}}
\put(540,1892){\makebox(0,0)[r]{0.15}}
\put(540,1345){\makebox(0,0)[r]{0.1}}
\put(540,798){\makebox(0,0)[r]{0.05}}
\put(540,251){\makebox(0,0)[r]{0}}
\end{picture}
\caption[dummy]{{\it The $E_{\rm cone}$ distribution as in \fig{Fig:cut}.
    Solid histogram is gluon fragmentation according to the octet
    model from \Ariadne\ to leading order on the parton level.  Dashed
    histograms are the same but with full evolution and hadronization:
    long-dashed with a delta function in the splitting function, and
    short-dashed a finite width of $\epsilon=v^2$. The dotted line is
    from an analytical calculation to leading order.}}
\label{Fig:ocut}
\end{figure}
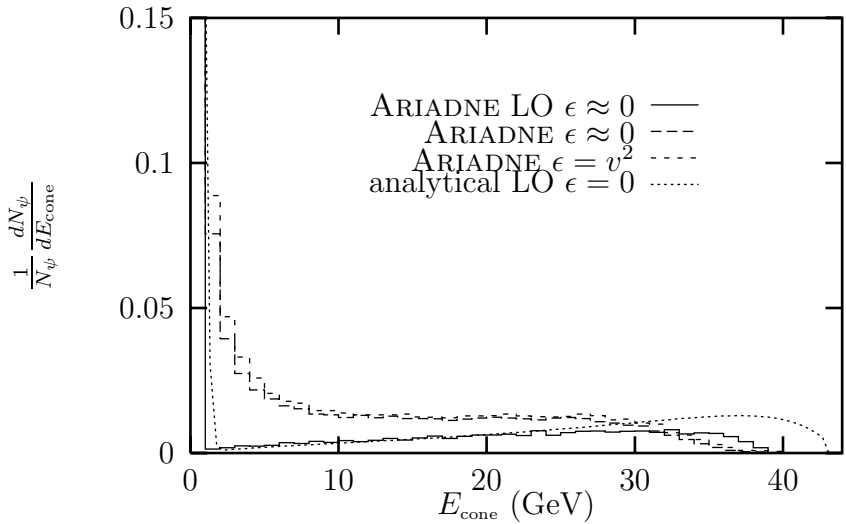

As an aside we want to point out that the threshold at $z=2
M_\psi/M_Z$ is reproduced in the analytic fragmentation formalism if
the distinction between the light cone momentum fraction $y = p_\psi^+
/ M_Z = (E_\psi + |{\bf p}_\psi|) / M_Z$ and the energy fraction $z =
2 E_\psi / M_Z = y + y^{-1} M_\psi^2/M_Z^2$ is taken into account.
This distinction has been neglected in equations such as
\eq{TwoTermFragmentation}, since fragmentation functions $D(y,\mu^2)$
are correct only up to corrections of \order{y^{-2}M_\psi^2/M_Z^2}.

\begin{figure}[t]
%\input /home/leif/MCEG/Projects/Onium/Z0paper/czcdist.tex
% GNUPLOT: LaTeX picture with Postscript
\setlength{\unitlength}{0.1bp}
\begin{picture}(3239,1943)(0,0)
\put(1582,1628){\makebox(0,0)[r]{Full \Ariadne}}
\put(1582,1728){\makebox(0,0)[r]{LO \Ariadne}}
\put(1828,51){\makebox(0,0){$z$}}
\put(100,1471){%
%
\makebox(0,0)[b]{\shortstack{$\frac{1}{N_{\mbox{\tiny had}}} \frac{dN_\psi}{dz}$}}%
%
}
\put(3056,151){\makebox(0,0){1}}
%\put(2810,151){\makebox(0,0){0.9}}
\put(2565,151){\makebox(0,0){0.8}}
%\put(2319,151){\makebox(0,0){0.7}}
\put(2074,151){\makebox(0,0){0.6}}
%\put(1828,151){\makebox(0,0){0.5}}
\put(1582,151){\makebox(0,0){0.4}}
%\put(1337,151){\makebox(0,0){0.3}}
\put(1091,151){\makebox(0,0){0.2}}
%\put(846,151){\makebox(0,0){0.1}}
\put(600,151){\makebox(0,0){0}}
\put(540,1892){\makebox(0,0)[r]{$1.0\cdot 10^{-4}$}}
\put(540,1072){\makebox(0,0)[r]{$0.5\cdot 10^{-4}$}}
\put(540,251){\makebox(0,0)[r]{0}}
\end{picture}

\caption[dummy]{{\it The $z$ distribution of $J/\psi$ from charm
    fragmentation without (upper curves) and with (lower curves)
    isolation cut. Solid histograms are from \Ariadne\ in leading
    order, and dashed histograms are from \Ariadne\ with full
    evolution and hadronization. }}
\label{Fig:singletc}
\end{figure}
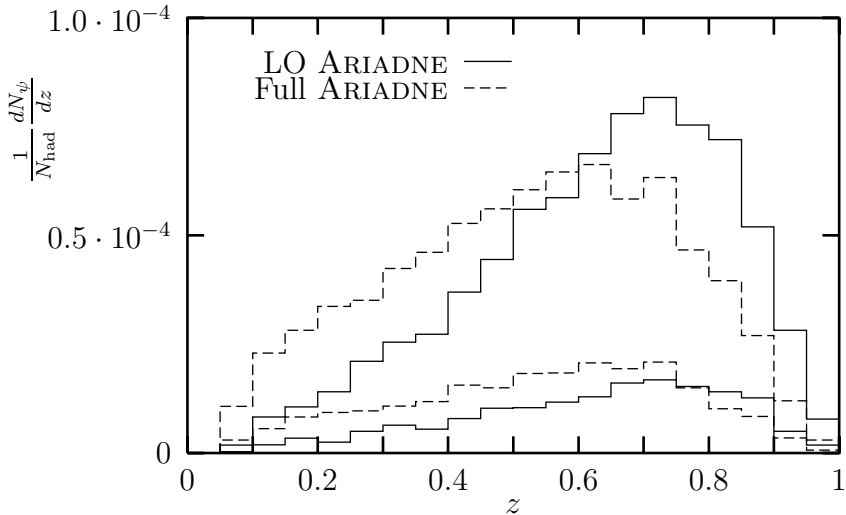

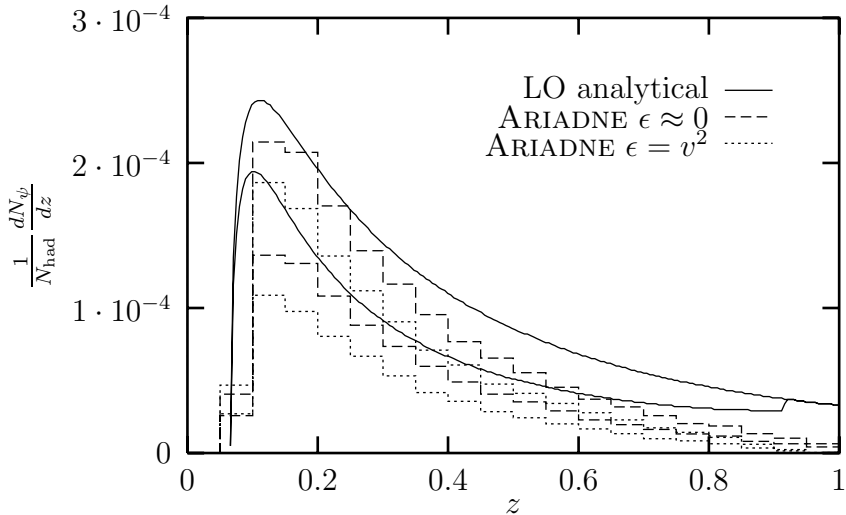
\begin{figure}[t]
%\input /home/leif/MCEG/Projects/Onium/Z0paper/ozcdist.tex
% GNUPLOT: LaTeX picture with Postscript
\setlength{\unitlength}{0.1bp}
\begin{picture}(3239,1943)(0,0)
\put(2565,1419){\makebox(0,0)[r]{\Ariadne\ $\epsilon=v^2$}}
\put(2565,1519){\makebox(0,0)[r]{\Ariadne\ $\epsilon\approx 0$}}
\put(2565,1619){\makebox(0,0)[r]{LO analytical}}
\put(1828,51){\makebox(0,0){$z$}}
\put(100,1071){%
%
\makebox(0,0)[b]{\shortstack{$\frac{1}{N_{\mbox{\tiny had}}} \frac{dN_\psi}{dz}$}}%
%
}
\put(3056,151){\makebox(0,0){1}}
\put(2565,151){\makebox(0,0){0.8}}
\put(2074,151){\makebox(0,0){0.6}}
\put(1582,151){\makebox(0,0){0.4}}
\put(1091,151){\makebox(0,0){0.2}}
\put(600,151){\makebox(0,0){0}}
\put(540,1892){\makebox(0,0)[r]{$3\cdot 10^{-4}$}}
\put(540,1345){\makebox(0,0)[r]{$2\cdot 10^{-4}$}}
\put(540,798){\makebox(0,0)[r]{$1\cdot 10^{-4}$}}
\put(540,251){\makebox(0,0)[r]{0}}
\end{picture}
\caption[dummy]{{\it The $z$ distribution of $J/\psi$ from
    colour-octet gluon fragmentation without (upper curves) and with
    (lower curves) isolation cut. Solid lines are from an analytical
    leading order calculation. Dashed histograms are from \Ariadne\ 
    with full evolution and hadronization: long-dashed with a delta
    function in the splitting function, and short-dashed a finite
    width of \mbox{$\epsilon=v^2$.}}}
\label{Fig:octetg}
\end{figure}

Our results for the direct prompt $J/\psi$ energy spectrum
$d\Gamma/dz$ before and after the cut on $E_{\rm cone}$, are presented
in Figs.\ \ref{Fig:singletc} and \ref{Fig:octetg}.  The shapes of the
distributions are not significantly changed by the evolution effects.

For the colour-singlet contribution from charm fragmentation (where
evolution effects can easily be included also in the analytical
calculation by means of differential equations, see Fig.\ 
\ref{Fig:Dc}), the total integrated branching ratio is ${\rm
  Br}^{(1)}(\ztodpj) = 0.27 \cdot 10^{-4}$, which is is the same as
the leading order result within the statistical error of 1\% in the
Monte Carlo calculation. This is consistent with the normal AP
evolution which does not change the overall normalization. Also the
improved evolution only differs from standard AP in the small-$z$
region, hence there is no large change in the normalization in that
case either. Note, however, that for the charm fragmentation we have
everywhere used a fixed $\alpha_s=0.253$ in the splitting function of
\eq{dc}. By default, \Ariadne\ would use a running $\alpha_s$ also
here, and with the $\Lambda_{\rm QCD}=0.22$ GeV fitted from global
event shapes, the result is 20\% higher.

For the the colour-octet gluon fragmentation channel using
\eq{doubledelta} we get ${\rm Br}^{(8)}(\ztodpj) = 0.58 \cdot
10^{-4}$, which is 15\% below the fixed-order analytical result of
$0.68 \cdot 10^{-4}$. Introducing a finite energy of the soft gluons
in the octet mechanism lowers the value even further to ${\rm
  Br}^{(8)}(\ztodpj) = 0.46\cdot 10^{-4}$.  Note, however, that the
value used for $\element{J/\psi}$ is from a fit \cite{ChoLeibovich} to
Tevatron data neglecting the energy of the soft gluons. But
high-$p_\perp$ hadroproduction cross sections are approximately
proportional to the fifth moment \mbox{$\int dz \; z^4 \, D(z,\mu)$}
of the fragmentation function, so that a trigger bias is introduced
and the normalization becomes very sensitive to the exact form of the
fragmentation function at large $z$. It turns out that changing from a
delta function in \eq{doubledelta} to a finite width
$\epsilon=0.2\approx v^2$ implies that $\element{J/\psi}$ has to be
doubled to give the same cross section at the Tevatron. This would
mean that the branching ratio becomes ${\rm Br}(Z^0 \to {\rm direct \;
  prompt}\; J/\psi + X) = 0.92\cdot 10^{-4}$ in the finite width case.
Also for the octet channel, the default running $\alpha_s$ in the
splitting function used in \Ariadne\ gives a 20\% higher branching
ratio than the fixed $\alpha_s$ used here.

To compare with experimental branching ratio of $Z^0$ into total
prompt $J/\psi$, the colour-octet contribution has to be multiplied by
a factor of 2.1 and the colour-singlet contribution by a factor of 1.3
to take into account $\chi_c$ and $\psi'$ decays into $J/\psi$.  The
resulting branching ratios are $1.57 \cdot 10^{-4}$ for $\epsilon=0$
and $2.28 \cdot 10^{-4}$ for $\epsilon=0.2$, both consistent with the
OPAL measurement \cite{OPAL:psi} ${\rm Br}(Z^0 \to {\rm total \;
  prompt}\; J/\psi + X) = (1.9 \pm 0.7 \pm 0.5 \pm 0.5) \cdot
10^{-4}$.

\section{Discussion \label{discussion}}

We have studied evolution effects in $Z^0$ fragmentation into prompt
$J/\psi$ by means of a Monte Carlo simulation. We worked within the
framework of nonrelativistic QCD (NRQCD), where the production of
charmonium proceeds through either colour-octet or colour-singlet
$c\bar c$ intermediate states.

Evolution effects turn out to be relatively small, diminishing the
colour-octet contribution to the decay width by 15\% while
colour-singlet contribution is unchanged. The shape of the $J/\psi$
energy spectrum also turns out to be stable against the evolution
effects. Thus the conclusion drawn by previous authors \cite{CKY,Cho}
on the basis of a leading-order calculation -- that the colour-octet
contribution has larger normalization and a much softer energy
spectrum than the colour-singlet contribution -- remains valid after
the inclusion of the evolution effects.

In fact, the dominant uncertainties in the comparison between NRQCD
predictions and experimentally measured branching ratios are due to
uncertainties in the fits of NRQCD matrix elements and to the small
statistics of the experimental results.

The Tevatron fit of the $\element{J/\psi}$ matrix element is subject
to a large uncertainty due to the trigger bias effect. The existing
fits of Tevatron data neglect the energy of the two soft gluons
emitted in the transition $c\bar c \left[ ^3S_1^{(8)} \right] \to \psi
+ gg$, thus maximizing the hardness of the fragmentation function.  A
realistic emitted energy of the order of $M_\psi v^2$ corresponds to a
softening of the fragmentation function which suppresses the cross
section by approximately a factor of two.  The fit value of the matrix
element would then have to be increased by this factor of two.

Apart from this and other theoretical uncertainties, the statistical
errors of the Tevatron data imply uncertainties in the matrix elements
around 30\%~\cite{ChoLeibovich}.

In the bottomonium sector, the approximation of a heavy quark mass is
better than in the charmonium sector. On the other hand, there is less
data available, and theoretical expressions involve more free
parameters than in the charmonium sector. In the Tevatron experiments,
contributions from $\chi_b$ decays are not resolved from direct
$\Upsilon$ production; in $Z^0$ decays, even the different radial
excitations of the $\Upsilon$ can currently not be resolved from each
other. In principle, however, the analysis can be easily extended to
the bottomonium sector \cite{Cho}.

The predictions of NRQCD are consistent with the observed quarkonium
production in $Z^0$ decays. However, both theoretical and experimental
uncertainties are large. A precision test of NRQCD may become possible
if uncertainties in the Tevatron fits of NRQCD matrix elements are
reduced and if more data become available at the $Z^0$ pole. Evolution
effects in the theoretical calculations are smaller than these
uncertainties and are under control, as we have shown above.

\end{document}